\documentclass[aps,preprint,floatfix,nofootinbib,showpacs,superscriptaddress]{revtex4-1}
\pdfoutput=1

\catcode`\@=11
\def\lsim{\mathrel{\mathpalette\@versim<}}
\def\gsim{\mathrel{\mathpalette\@versim>}}
\def\@versim#1#2{\vcenter{\offinterlineskip
\ialign{$\m@th#1\hfil##\hfil$\crcr#2\crcr\sim\crcr } }}
\catcode`\@=12

\catcode`@=12
\usepackage[export]{adjustbox}	
\usepackage{float}
\usepackage{epsfig,bbm}
\usepackage{amsmath}
\usepackage{slashed}
\usepackage{tikz}
\usepackage{bm}
\usepackage{amssymb}
\usepackage{mathtools}
\usepackage[caption=false]{subfig}
\usepackage{graphicx}
\usepackage{hyperref}
\usepackage{cleveref}

\def\lsim{\mathrel{\rlap{\lower4pt\hbox{\hskip1pt$\sim$}}
    \raise1pt\hbox{$<$}}} 
\def\gsim{\mathrel{\rlap{\lower4pt\hbox{\hskip1pt$\sim$}}
    \raise1pt\hbox{$>$}}} 

\newcommand{\newc}{\newcommand}
\newc{\renewc}{\renewcommand}
\newc{\red}{\textcolor{red}}
\newc{\blue}{\textcolor{blue}}

\begin{document}
\thispagestyle{empty}

\title{\Large Leptonic New Force and Cosmic-ray Boosted Dark Matter for the XENON1T Excess}

\author{Yongsoo Jho}
\email{jys34@yonsei.ac.kr}
\affiliation{Department of Physics and IPAP, Yonsei University, \\
Seoul 03722, Republic of Korea}

\author{Jong-Chul Park}
 \thanks{co-corresponding author}
\email{jcpark@cnu.ac.kr}
\affiliation{Department of Physics and Institute of Quantum Systems (IQS), Chungnam National University, Daejeon 34134, Republic of Korea}

\author{Seong Chan Park}
 \thanks{co-corresponding author}
\email{sc.park@yonsei.ac.kr}
\affiliation{Department of Physics and IPAP, Yonsei University, \\
Seoul 03722, Republic of Korea}

\author{and Po-Yan Tseng}
\email{tpoyan1209@gmail.com}
\affiliation{Department of Physics and IPAP, Yonsei University, \\
Seoul 03722, Republic of Korea}


\preprint{YHEP-COS20-06}

\begin{abstract}
The recently reported excess in XENON1T is explained  by new leptonic forces, which are free from gauge anomalies.
We focus on two scenarios with and without dark matter.
In Scenario \#1, the gauge boson of gauged lepton number U(1)$_{L_e-L_j}$, $j=\mu$ or $\tau$  provides non-standard interaction between solar neutrino and electron that enhances the number of electron recoil events in the XENON1T detector.
In Scenarino \#2, the new gauge boson exclusively couples to electron and dark matter, then cosmic-ray electrons can transfer their momenta to dark matter in halo.
The boosted dark matter generates the electron recoil signals of ${\cal O}(1)$ keV.
The dark matter, aided by the new gauge interaction, efficiently heats up a neutron star in our Galaxy more than $\sim1500$ K as a neutron star captures the halo dark matter.
Therefore, we propose to utilize the future infrared telescope to test our scenario.
\end{abstract}

\maketitle


\section{Introduction}

An excess in low energy electronic recoil events over known backgrounds has been reported by the XENON1T collaboration~\cite{Aprile:2020tmw}.
The excess is rising towards lower energies below 7 keV and most prominent in 2-3 keV, which may be simply due to the Tritium~\cite{Aprile:2020tmw} or Ar-37~\cite{Szydagis:2020isq} contamination.
New particles beyond the standard model (SM) such as solar axions, solar neutrinos, light axion-like-particles (ALPs), and dark photon dark matter have been proposed to explain the excess by the XENON1T collaboration~\cite{Aprile:2020tmw}, but none of them is favored by experimental and observational data~\cite{Viaux:2013lha,Ayala:2014pea,Bertolami:2014wua,Battich:2016htm,Giannotti:2017hny, Corsico:2014mpa,Diaz:2019kim}.
The XENON1T result prefers the electron recoil spectrum by boosted dark matter (DM)~\cite{Kannike:2020agf}\footnote{The authors of Ref.~\cite{Giudice:2017zke}, for the first time, pointed out that the XENON1T experiment can be sensitive to fast-moving or boosted dark matter interacting with electrons.},
and there exist various suggested mechanisms, for example by cosmic-ray~\cite{Su:2020zny,Cao:2020bwd}, Sun~\cite{Chen:2020gcl}, decay or annihilation~\cite{Fornal:2020npv,Du:2020ybt,Alhazmi:2020fju,Choi:2020udy,Buch:2020mrg} of heavier particles.
It can also be explained by non-standard neutrino-electron interactions coming from neutrino magnetic moment~\cite{Khan:2020vaf} or dark gauge boson~\cite{Boehm:2020ltd,Bally:2020yid,AristizabalSierra:2020edu}.
Inelastic DM scenarios are discussed in Refs.~\cite{Bell:2020bes,Paz:2020pbc}.
In addition, dark photon dark matter~\cite{Alonso-Alvarez:2020cdv,Nakayama:2020ikz}, solar axion~\cite{DiLuzio:2020jjp}, and Migdal effect~\cite{Dey:2020sai} have been studied.

In this work, we propose two scenarios with anomaly free, leptonic gauge symmetries:
\begin{itemize}
\item{Scenario \#1:}
The solar neutrinos can interact with electrons in the XENON1T detector by the exchange of the new gauge boson $X$ and generate signals~\cite{Harnik:2012ni}.
The gauged symmetry U(1)$_{L_e-L_i}$ with $i=\mu$ or $\tau$ is chosen\footnote{In general, the gauge symmetries of lepton numbers and baryon numbers in the form $(L_i-L_j) + \epsilon(B_k-L_k)$ with various combinations of different generations ($i,j,k,=1,2,3$) are anomaly free~\cite{Huh:2013hga,Bauer:2018onh}.
Also see \cite{Jho:2020jsa} for the recent J-PARC KOTO anomaly.}  then the Lagrangian includes the interactions of the new gauge boson $X$ with electron and neutrino:
\begin{align}
\label{eq:1}
{\cal L}_{Scen\#1} &\supset -X_\mu \left(g_e J^\mu_e +g_\nu J^\mu_\nu\right) +\cdots\,,
\end{align}
where ``$\cdots$'' includes the kinetic terms and also interactions with other leptons.
The electronic and neutrino currents are respectively given as
\begin{align}
J_e^\mu = \bar{e}\gamma^\mu e, \quad
J_\nu^\mu = \bar{\nu_L}\gamma^\mu \nu_L.
\end{align}
\item{Scenario \#2:}
The dark matter, as well as electron, couples with the new gauge boson $X$.
The DM particles boosted by cosmic-ray electrons~\cite{Bringmann:2018cvk,Ema:2018bih,Cappiello:2019qsw,Dent:2019krz} can interact with electrons in the XENON1T detector and produce the observed excessive signals.
\begin{align}
\label{eq:3}
{\cal L}_{Scen\#2} &\supset -X_\mu \left(g_e J^\mu_e + g_\chi J^\mu_\chi \right)+\cdots\,.
\end{align}
To avoid the sizable stellar cooling effect by neutrinos, we turn off the neutrino couplings.
Indeed, this assumption ensures our setup anomaly free.
The dark matter current for Dirac spinor $\chi$ is given as
\begin{align}
J_\chi^\mu = \bar{\chi}\gamma^\mu \chi\,.
\end{align}
\end{itemize}

\section{Scenario \#1}

\begin{figure}[t]
\centering
\includegraphics[width=.8\textwidth]{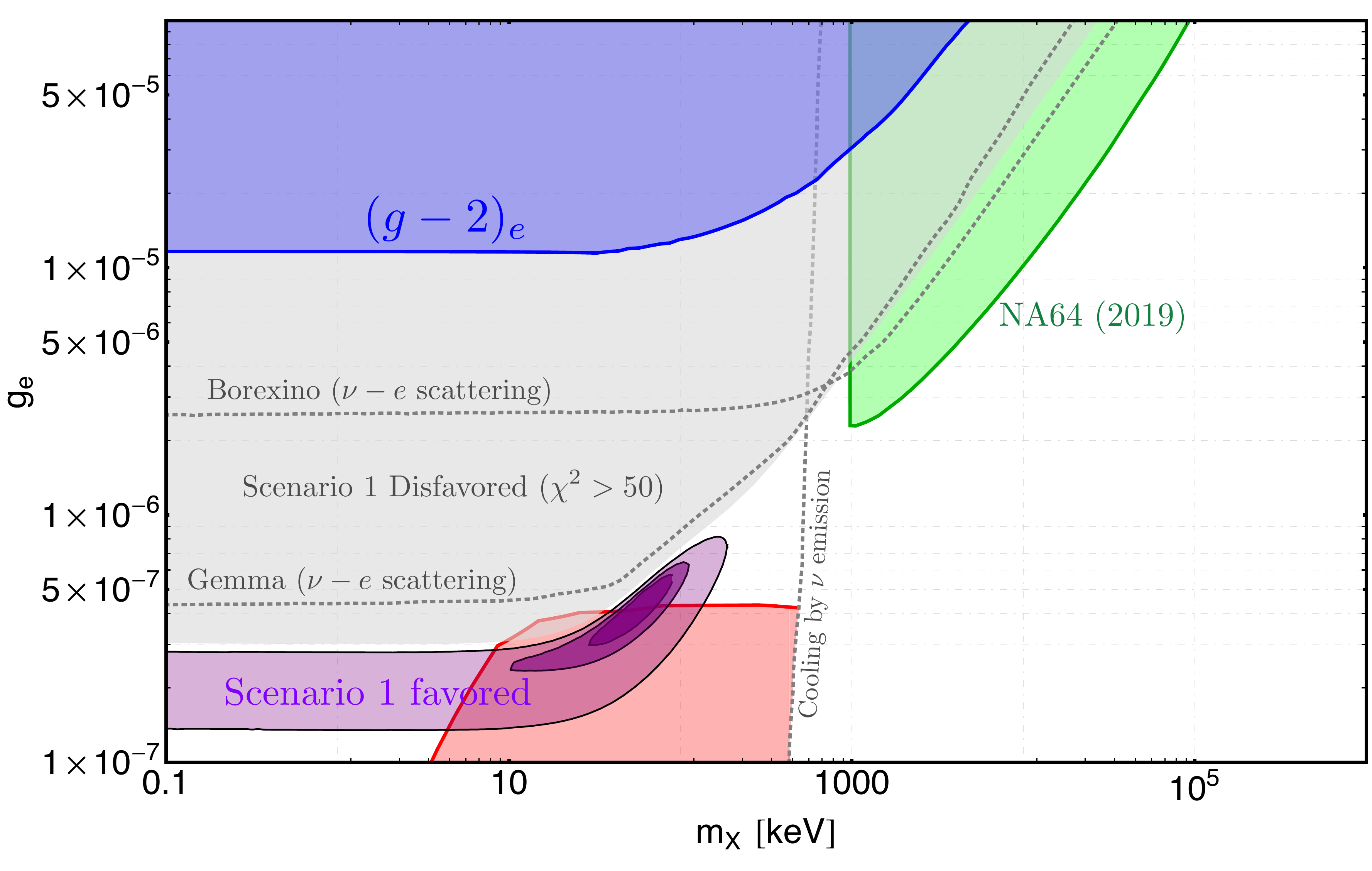}
\caption{\small \label{fig_scenario1_constraints}
[Scenario \#1] The dark purple and light purple regions are favored by the XENON1T data.
The best-fit is $(m_X, g_e)=(45.5\,{\rm keV}, 3.2\times 10^{-7})$ with $\chi^2_{\rm best-fit}=42.03$ and $p$-value=0.109, compare to the background model $B_0$ with $\chi^2_{\rm B_0}=46.49$.
We show the contours of $\chi^2 = 43.0, 44.0, 46.0$ as the boundaries of purple shaded regions.
The gray shaded region is disfavored as $\chi^2 > 50$.
The constraints on electron coupling $g_e$ from $(g-2)_e$ (blue), the cooling of the Sun and Gloubular clusters (red), and the missing momentum search for invisibly decaying dark photon by the NA64 experiment (green) are also shown.
The gray dotted lines corresponds to the constraints in the presence of $g_e$ and $g_\nu$, from $\nu$-$e$ scattering at Borexino and Gemma and the stellar cooling by neutrino emissions.
}
\end{figure}

In Scenario \#1, we consider the electron recoil spectrum from the non-standard scattering between the solar neutrino and the electron in the XENON1T detector via the exchange of a $X$ boson from Eq.(\ref{eq:1}).
The solar neutrinos are produced by main processes of solar nuclear reaction chains~\cite{Bahcall:1964gx}.
In the range of keV--MeV, these processes can generate sizeable amount of neutrinos which can give raise to keV electron recoils and amplify signals at the XENON1T detector via the $X$ boson exchange.
The energy spectrum of the total recoil event rate is given by
\begin{eqnarray}
\frac{dR^X}{dK_r} & = & N_T \cdot \epsilon (K_r) \cdot \int_{K_\nu^{\min}(K_r)}^\infty \frac{d\Phi_{\text{solar }\nu}}{dK_\nu} \frac{d\sigma_{\nu e}^X}{dK_r} dK_\nu\,,
\end{eqnarray}
where $K_\nu$ and $K_r$ are respectively the kinetic energy of the solar neutrino and the electron recoil energy, $\epsilon (E_r)$ is the efficiency of electron recoil in the XENON1T detector.
For the total number of target electrons $N_T$, we conservatively take into account 26 electrons in $5p$ to $4s$ atomic shells from each Xe atom~\cite{Cao:2020bwd}, because the electron in inner shells has a binding energy larger than $\mathcal{O}$(keV).
The energy resolution of detector is taken into account
by convolution with the resolution function~\cite{Bramante:2020zos}.
In the presence of a new leptophilic U(1)$_{L_e - L_i}$,
the gauge boson $X$ couples to electron and neutrinos with equal strength $g_e=g_\nu$, and the differential cross section of neutrino-electron scattering $\nu e^- \to \nu e^-$ is given by~\cite{Cao:2020bwd}
\begin{eqnarray}
\frac{d\sigma_{\nu e}^X}{dK_r} =
\frac{(g_e g_\nu)^2}{4\pi}
\frac{2m_e(m_\nu+K_\nu)^2-K_e\left\{(m_\nu+m_e)^2+2m_eK_\nu\right\}+m_eK^2_e}
{(2m_\nu K_\nu+K^2_\nu)(2m_e K_e+m^2_X)^2}\,.
\end{eqnarray}
Here for a given recoil energy $K_r$, the required minimum value of the neutrino energy is
\begin{eqnarray}
K_\nu^{\min} & = & \frac{1}{2} \left(K_r + \sqrt{K_r^2 + 2 K_r m_e} \right)\,.
\end{eqnarray}
For the consistency check with the XENON1T data~\cite{Aprile:2020tmw}, we perform the $\chi^2$ minimization for the electron recoil spectrum $dR/dK_r$ by summing up the background model ($B_0$) and the new physics contribution.
In Fig.~\ref{fig_scenario1_constraints} the purple shaded region shows the favored region, and the best-fit point corresponds to $(m_X, g_e)=(45.5\,{\rm keV}, 3.2\times 10^{-7})$ with $\chi^2_{\rm best-fit}=42.03$ and $p$-value=0.109, which means that the Scenario \#1 hypothesis is favored over $B_0$ at $1.64\,\sigma$~\cite{Cheung:2018ave} by the data.

However, the $X$ boson in the parameter region favored by the XENON1T excess can significantly speed up the cooling process of the Sun and Globular clusters because the $X$ boson with $m_X\simeq 100\,{\rm keV}$ can be copiously produced by thermal radiation of electrons and the solar plasmon resonance.
If the coupling $g_e$ of $X$ is small enough, $X$ can escape from the Sun and contribute extra cooling, which is constrained as ``Stellar Cooling'' \cite{Harnik:2012ni} and shown as the red shaded region.
The favored region of the U(1)$_{L_e-L_i}$ $X$ boson model for the XENON1T excess is significantly constrained by the stellar cooling.
Duo to the non-trivial couplings of $g_e$ and $g_\nu$, even if the mass of $X$ is outside the plasmon resonance region, neutrinos can be still produced through an off-shell $X$ and escape the Sun.
On the other hand, when the coupling is large enough, i.e., $g_e \gtrsim 4\times 10^{-7}$, the $X$ boson cannot escape the Sun.
However, its decay into neutrinos still contributes to the cooling.
Thus, these relevant parameter regions are excluded and called ``Cooling via $\nu$ emission'' \cite{Harnik:2012ni, Davidson:2000hf}.
We also show other constraints from $\nu$-$e$ scattering~\cite{Bellini:2011rx, Beda:2010hk}, $(g-2)_e$~\cite{Pospelov:2008zw}, and NA64~\cite{NA64:2019imj} in Fig.\ref{fig_scenario1_constraints}.
For more details, see Section~\ref{sec:constraints}.

\section{Scenario \#2}

In this section, we consider two possibilities of the interaction between DM and electron, one is to use simply the effective DM-electron scattering cross section $\sigma_{\rm DM-e}$, and the other comes from
the exchange of the $X$ gauge boson as describe in Eq.~(\ref{eq:3}).
For the later case, we adopt the differential cross section for the process ${\rm DM} e \to {\rm DM} e$
as a function of the recoil energy of the target electron $K_e$~\cite{Cao:2020bwd}:
\begin{equation}
\frac{d\sigma_X({\rm DM}e \to {\rm DM}e)}{dK_e}=
\frac{(g_e g_{\chi})^2}{4\pi}\,
\frac{2m_e(m_{\rm DM}+K_{\rm DM})^2
      -K_e\left\{(m_e+m_{\rm DM})^2+2m_e K_{\rm DM}\right\}
      +m_e K^2_e}
{(2m_{\rm DM}K_{\rm DM}+K^2_{\rm DM})(2m_e K_e +m^2_X)^2}
\,,
\end{equation}
where $K_{\rm DM}$ is the DM kinetic energy.
The maximum electron recoil energy is given by
\begin{equation}
K^{\rm max}_e(K_{\rm DM}) = \frac{2 m_e(K^2_{\rm DM}+2m_{\rm DM}K_{\rm DM})}
{(m_{\rm DM}+m_e)^2+2m_e K_{\rm DM}}\,.
\end{equation}

Due to the DM-electron interaction, the non-relativistic halo DM will be boosted by high energy cosmic-ray electrons $d\Phi_e /d\Omega$,
which are from the observations of AMS-02, DAMPE, Fermi-LAT, and Voyager in the range of $2\,{\rm MeV}\,\leq K_e \leq 90\,{\rm GeV}$~\cite{Boschini:2018zdv}.
The boosted DM flux can be obtained by convolution of the cosmic electron flux and DM-electron differential cross section~\cite{Ema:2018bih}
\begin{eqnarray}
\frac{d\Phi_{\rm DM}}{d\Omega}(K_{\rm DM},b,l)=
\frac{J(b,l)}{m_{\rm DM}}\int dK_e \frac{d\Phi_e}{d\Omega}
\frac{d\sigma_{{\rm DM}e \to {\rm DM}e}}{dK_{\rm DM}}\,,
\end{eqnarray}
where $J(b,l)=\int_{l.o.s}d\ell \rho_{\rm DM}$
is the line of sight integral of the DM mass density $\rho_{\rm DM}$ along the direction of galactic coordinates $(b,l)$.
For the effective cross section case, we simply replace the differential cross section
$d\sigma_{{\rm DM}e \to {\rm DM}e}/dK_{\rm DM}$
with $\sigma_{\rm DM-e}/K^{\rm max}_{\rm DM}(K_e)$
in the above equation.
Then we use the boosted DM flux to compute the
electron recoil energy spectrum for XENON1T.

\begin{figure}[t]
\centering
\includegraphics[height=2.9in,angle=270]{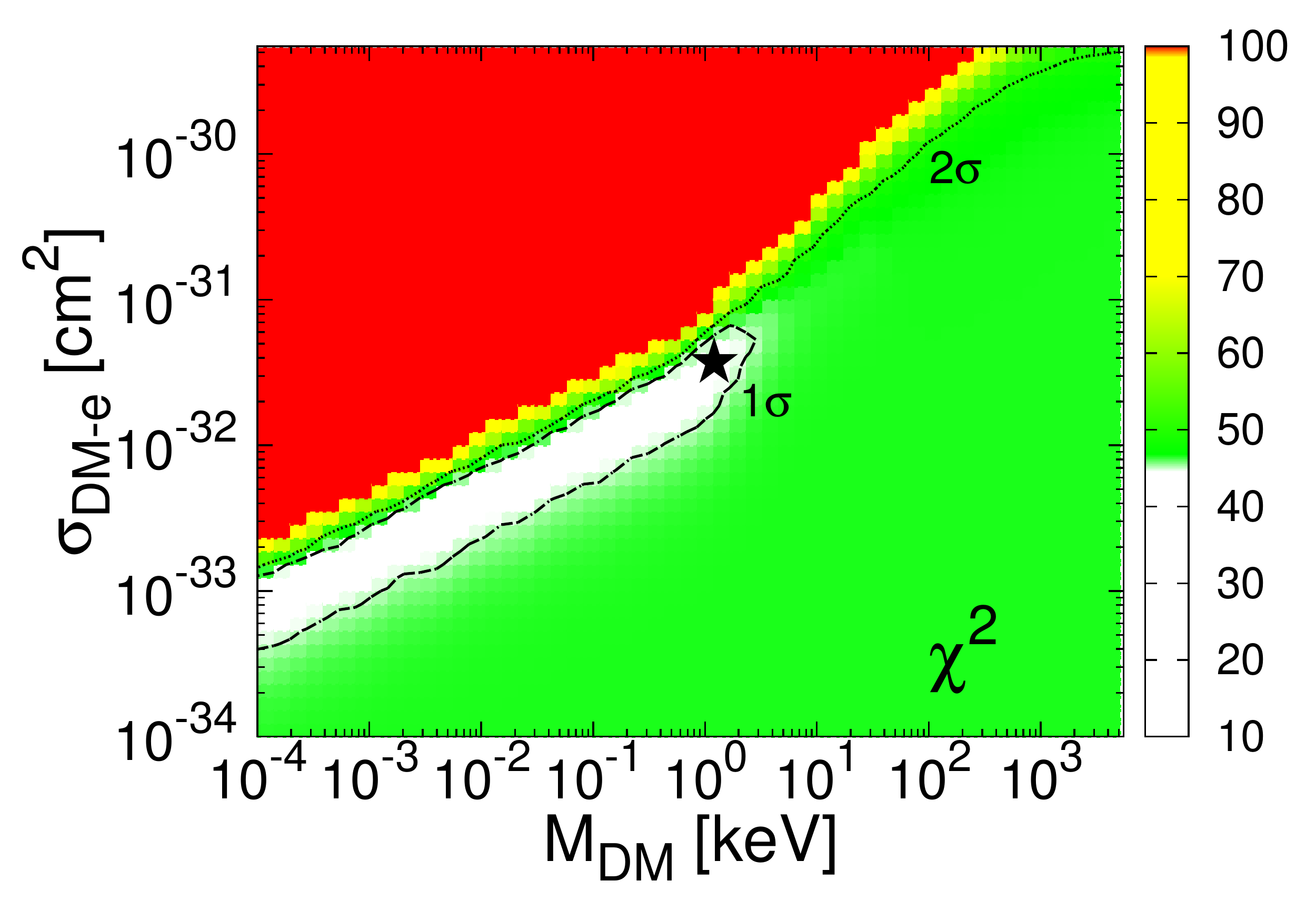}
\includegraphics[height=2.9in,angle=270]{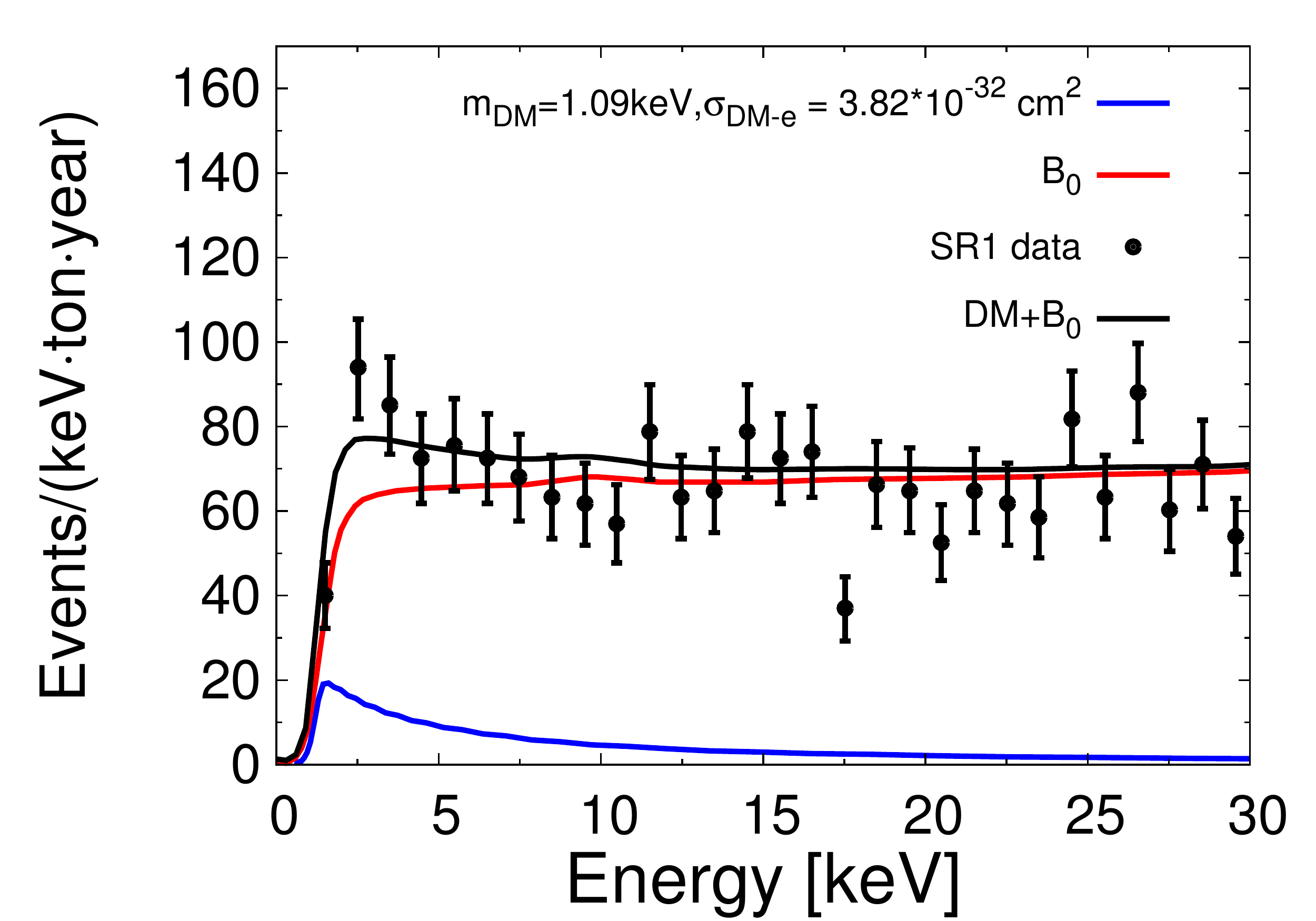}
\caption{\small \label{fig:effective_sigma_XENON1T}
[Scenario \#2: the effective DM-electron scattering cross section case]
Left-panel: The $\chi^2$ distribution in the  $(m_{\rm DM}, \sigma_{\rm DM-e})$ plane by fitting to the XENON1T event spectrum.
The best-fit point provides $\chi^2_{\rm best-ft}=42.82$ and $p-{\rm value}=0.160$ labelled with the ``star'' symbol,
where the curve $B_0$ gives $\chi^2_{B_0}=46.49$.
The 1$\sigma$ and 2$\sigma$ regions correspond to $\Delta \chi^2\equiv \chi^2-\chi^2_{\rm best-fit}=2.30$ and $5.99$, respectively.
Right-panel: The event spectrum for the best-fit point.
}
\end{figure}

In Fig.~\ref{fig:effective_sigma_XENON1T}, the $\chi^2$ fits to the XENON1T data are shown for the effective DM-electron cross section case.
The best-fit point $(m_{\rm DM}, \sigma_{\rm DM-e})=(1.09\,{\rm keV}, 3.82\times 10^{-32}{\rm cm^2})$ gives $\chi^2_{\rm best-fit}=42.82$, which is slightly better than the background model $B_0$ of $\chi^2_{\rm B_0}=46.49$.
As can be seen in the left-panel, the XENON1T excess prefers $m_{\rm DM}\lesssim 10$ keV and linear correlation between $\sigma_{\rm DM-e}$ and $m_{\rm DM}$.
Since the event rate is proportional to the square of $\sigma_{\rm DM-e}$, the upper-left corner (red region) overproduces the events, and hence is ruled out by the XENON1T results.
The ``star'' labels the best-fit point, and the corresponding spectrum is shown in the right-panel which features a peak around $2-3$ keV.

\begin{figure}[h]
\centering
\includegraphics[height=2.9in,angle=270]{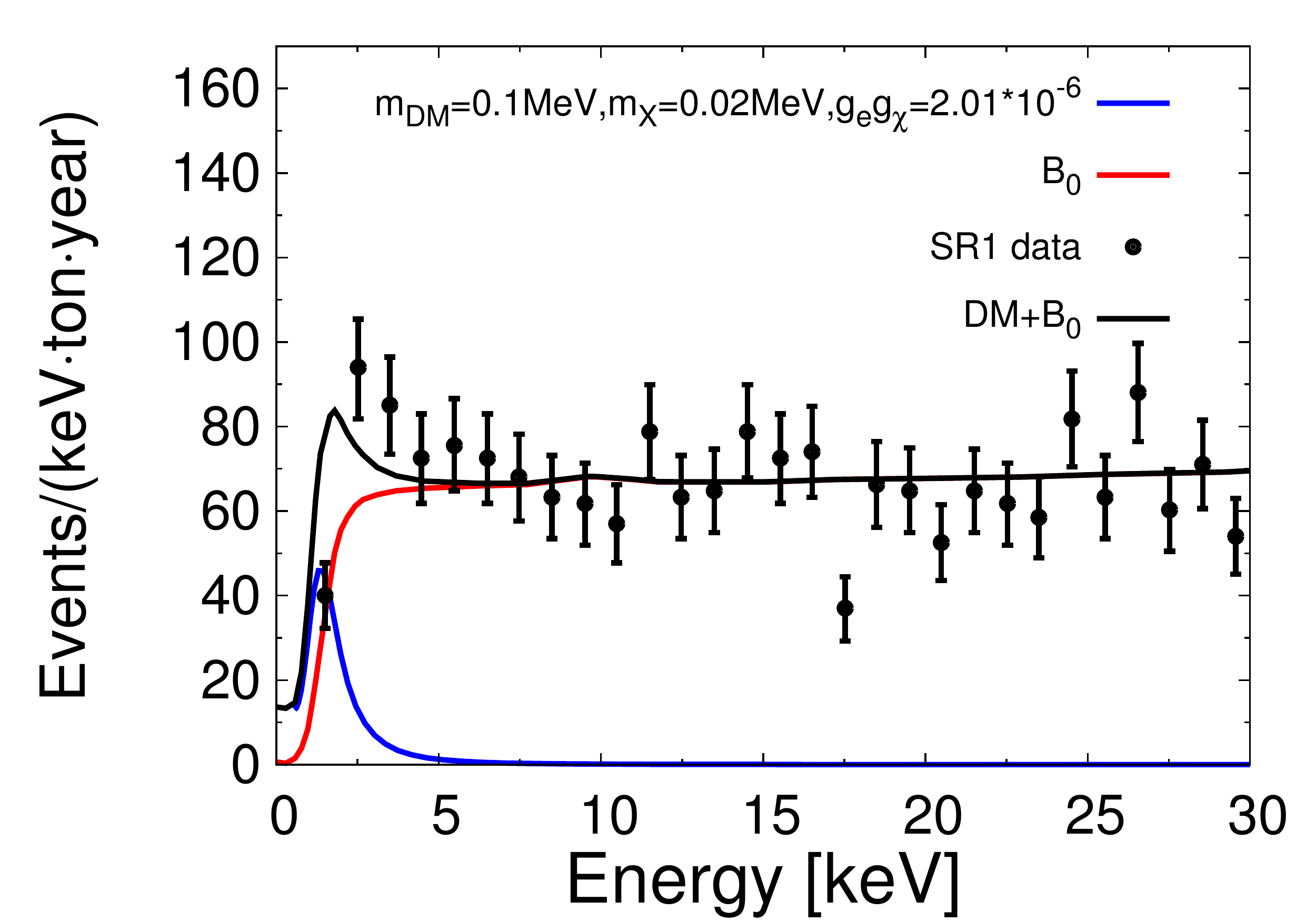}
\includegraphics[height=2.9in,angle=270]{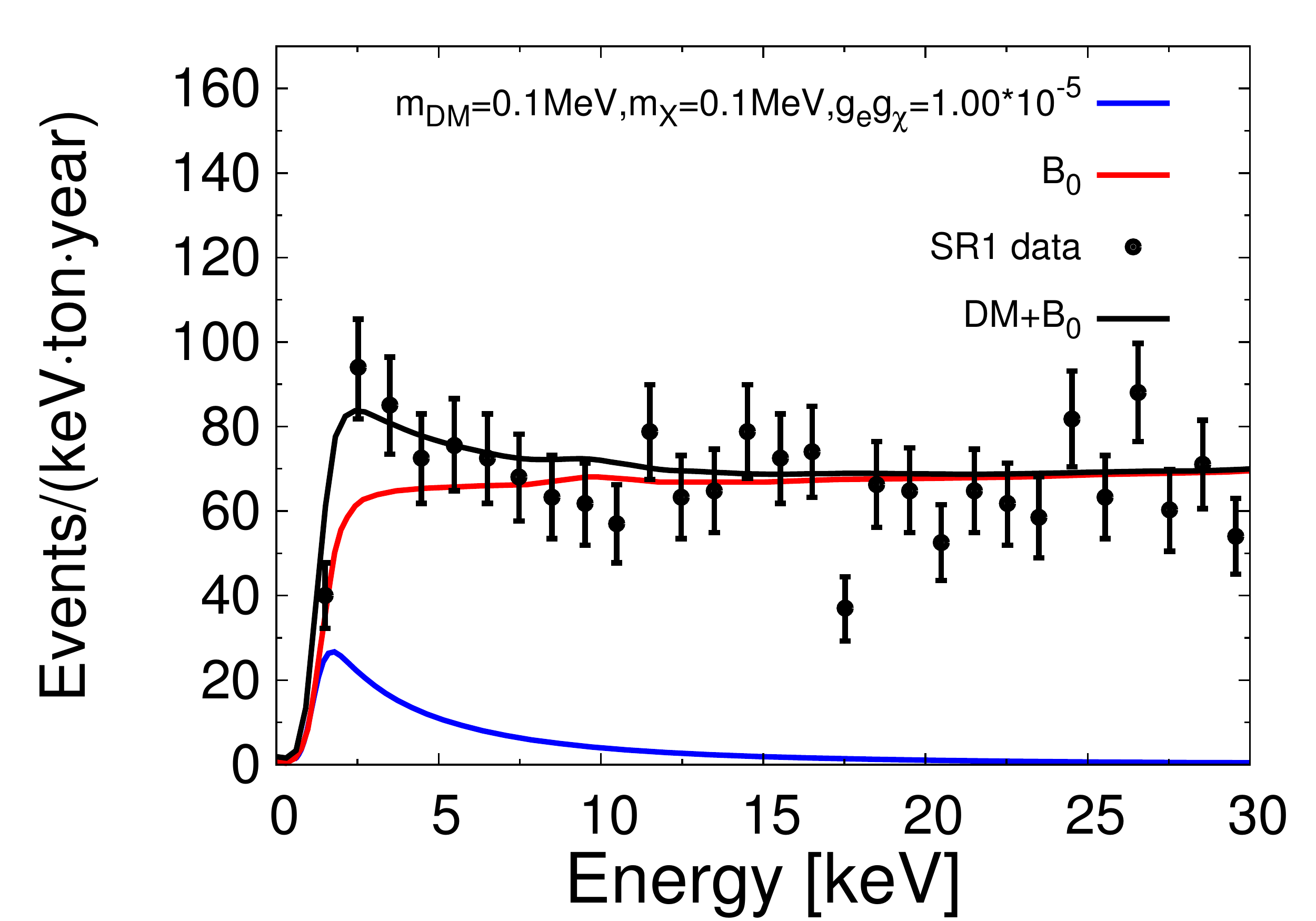}
\includegraphics[height=2.9in,angle=270]{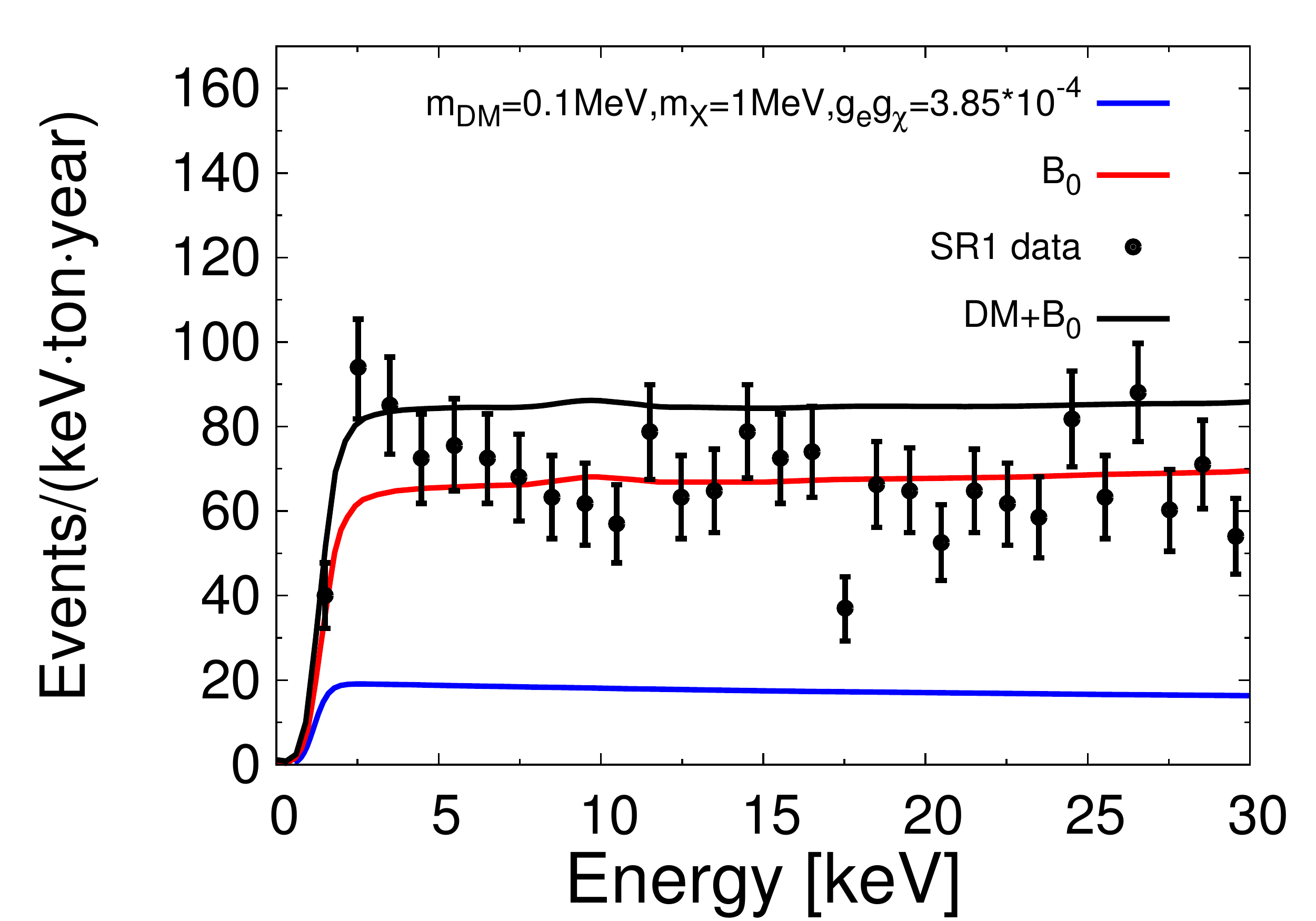}
\includegraphics[height=2.9in,angle=270]{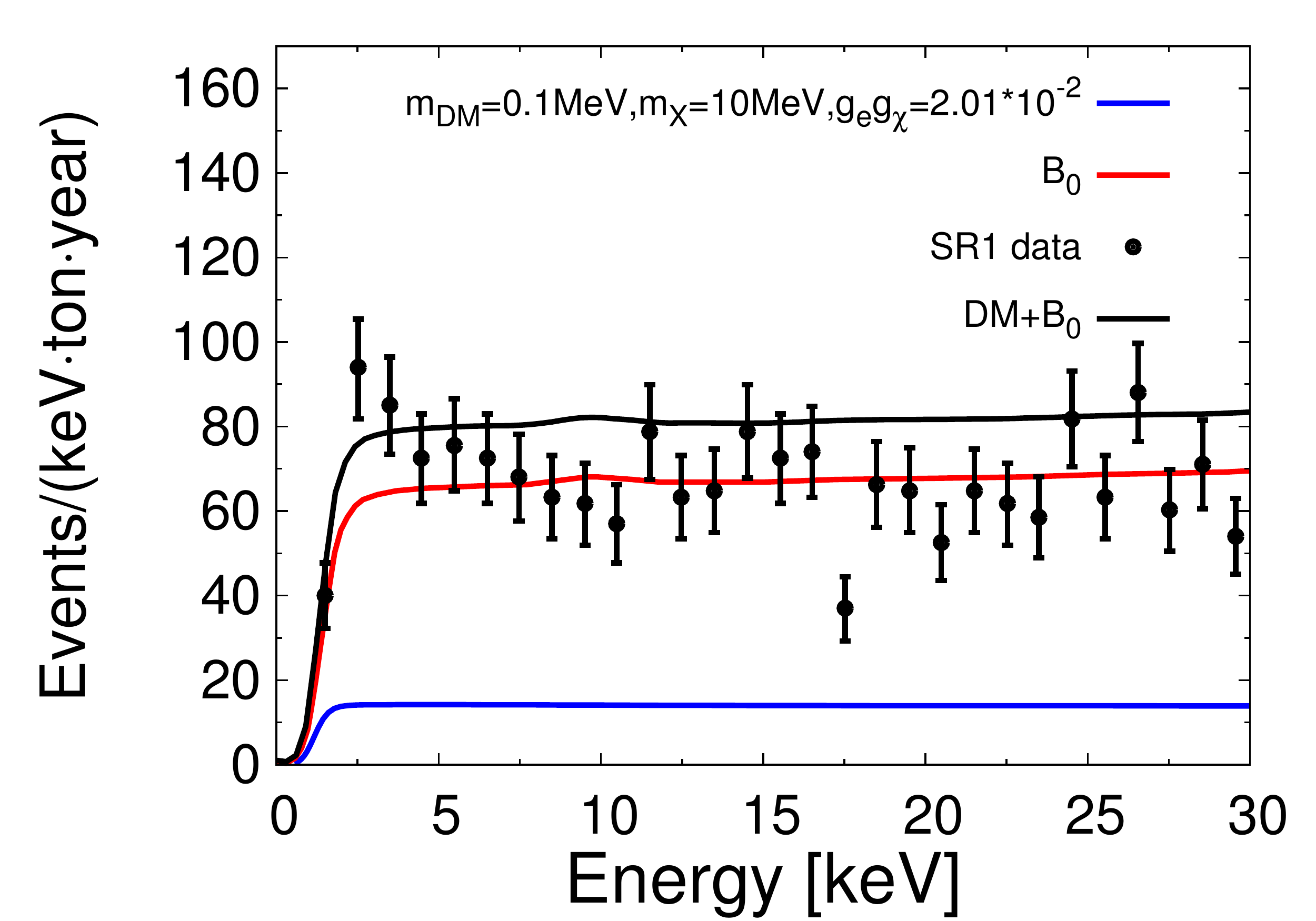}
\caption{\small \label{fig:DM_Xboson_XENON1T}
[Scenario \#2: the $X$ gauge boson mediator case]
The sample event spectra are shown for four benchmark values of $m_X=0.02,0.1,1$, and $10$ MeV assuming $m_{\rm DM}=0.1$ MeV.
}
\end{figure}
\begin{figure}[h]
\centering
\includegraphics[height=2.9in,angle=270]{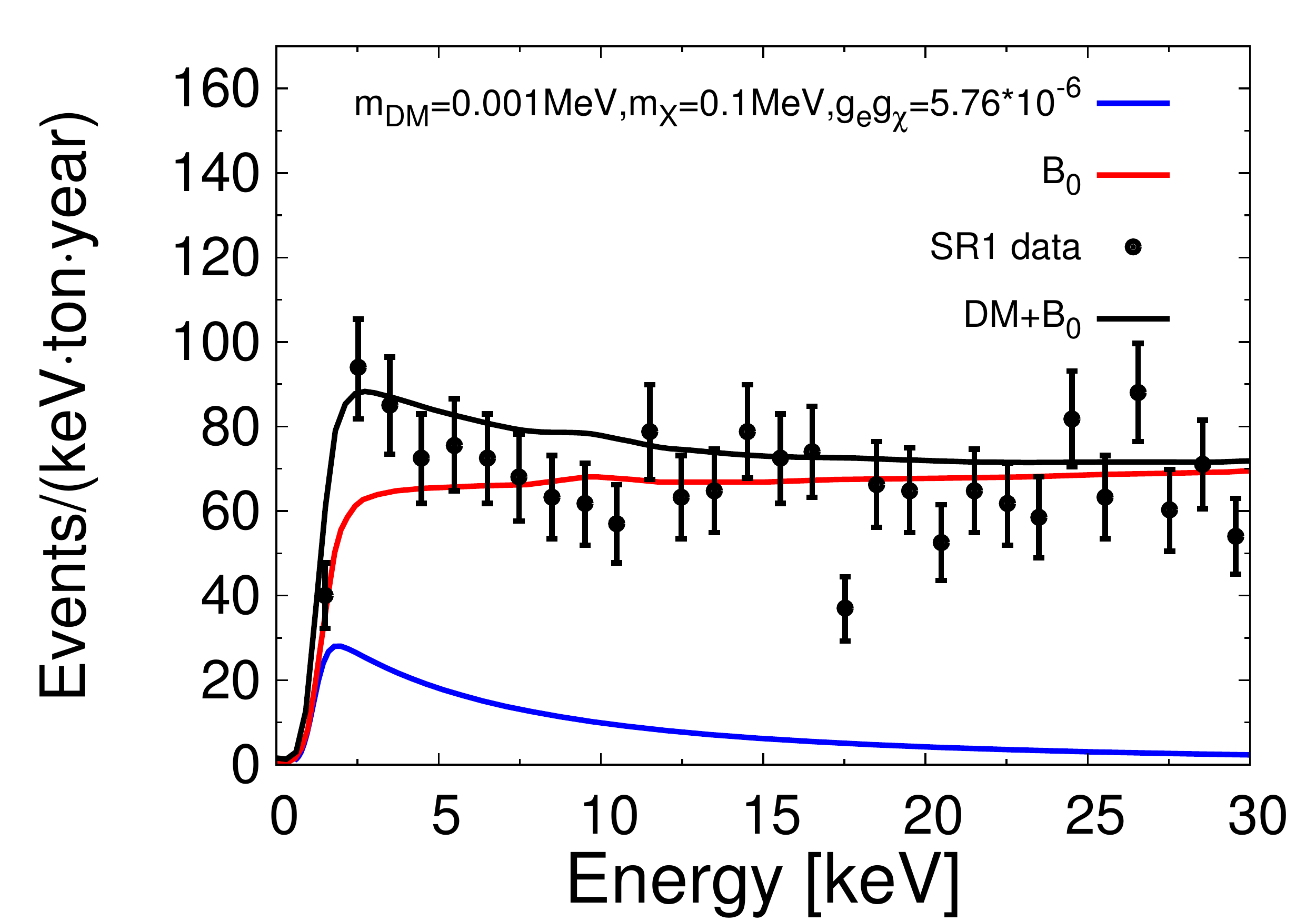}
\includegraphics[height=2.9in,angle=270]{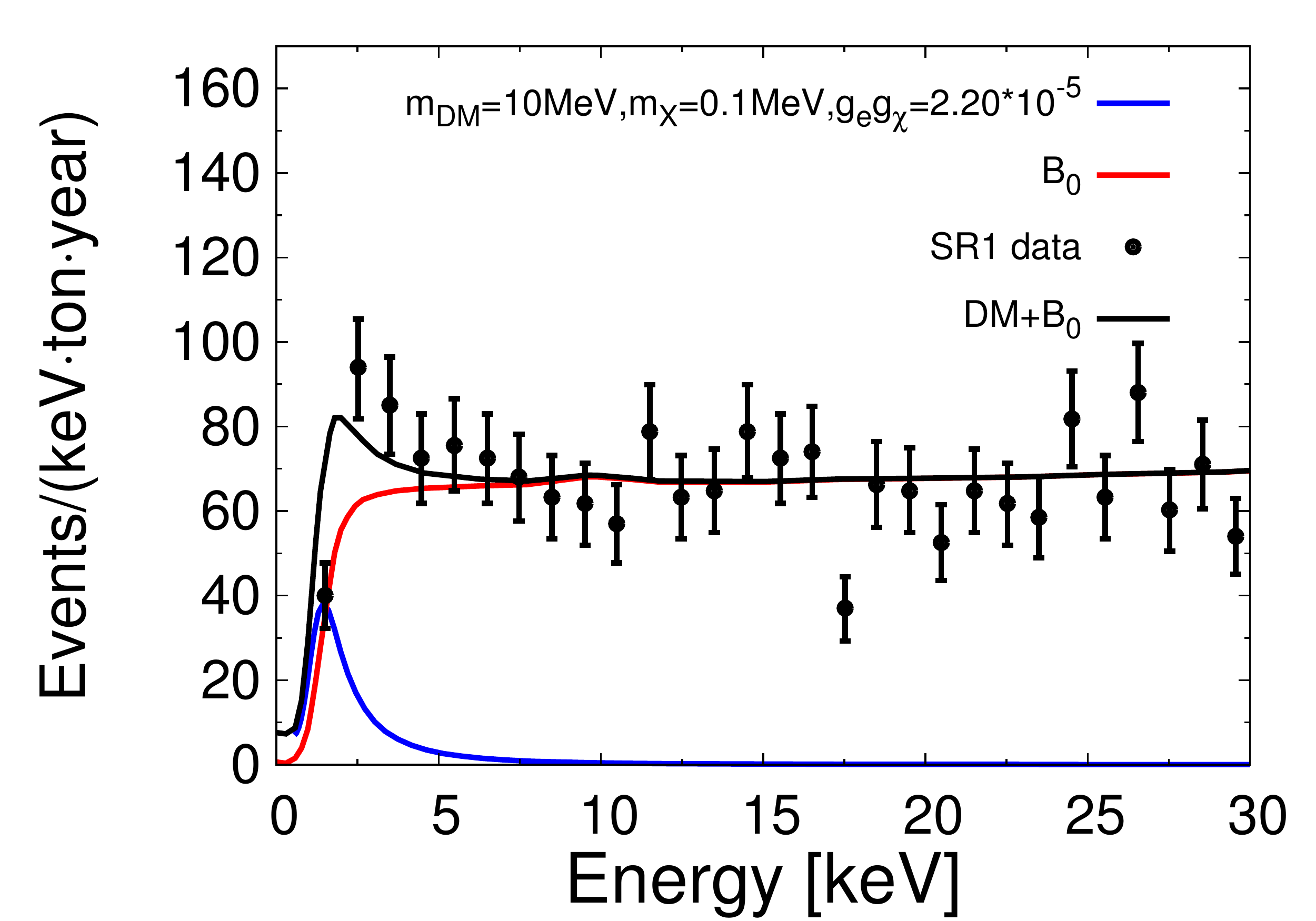}
\caption{\small \label{fig:DM_Xboson_XENON1T_2}
[Scenario \#2: the $X$ gauge boson mediator case]
The event spectra are shown for two representative values of $m_{\rm DM}=0.001$ and $10$ MeV with a fixed $m_X=0.1$ MeV.
}
\end{figure}

\begin{figure}[h]
\centering
\includegraphics[height=2.9in,angle=270]{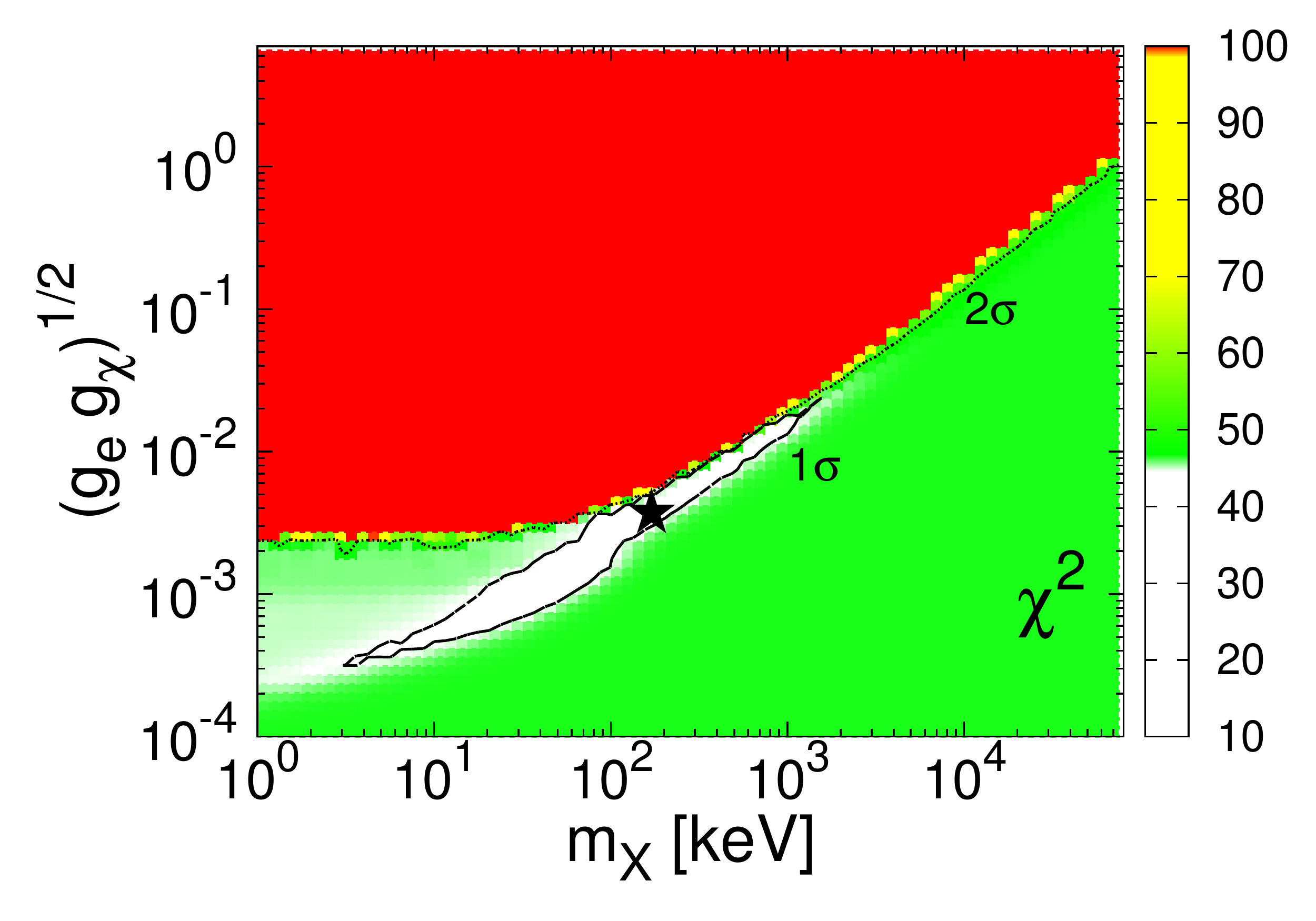}
\includegraphics[height=2.9in,angle=270]{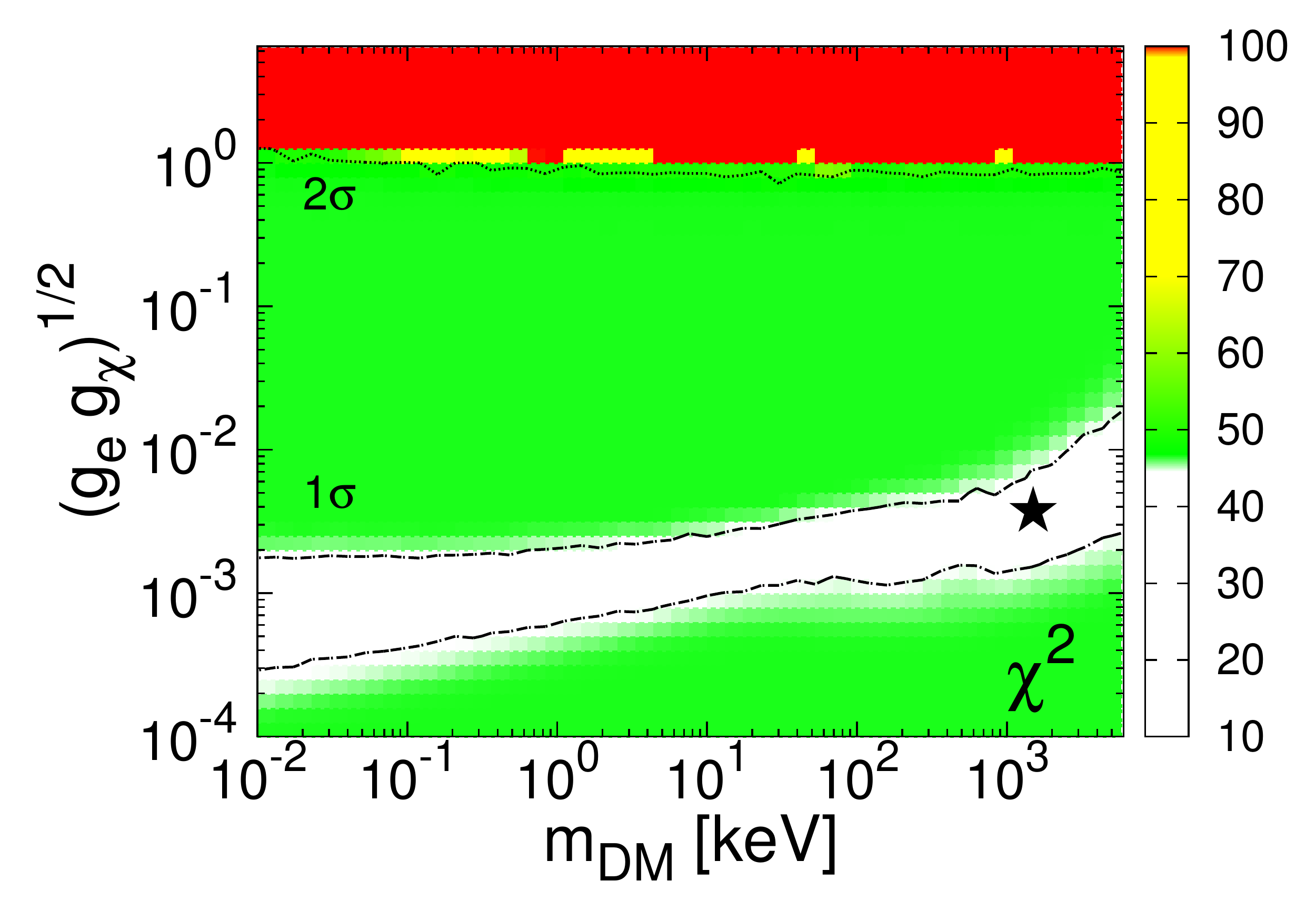}
\includegraphics[height=2.9in,angle=270]{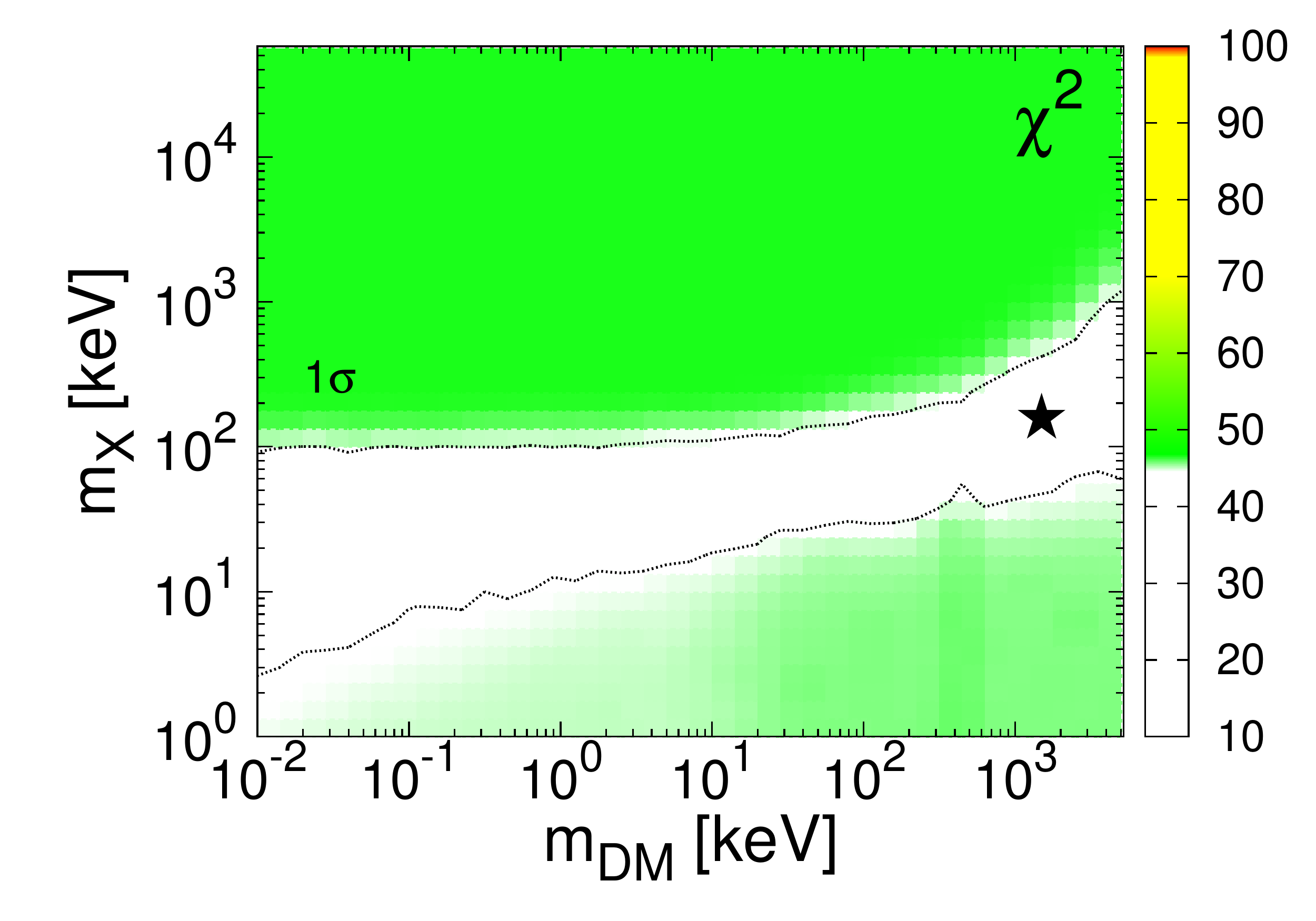}
\includegraphics[height=2.9in,angle=270]{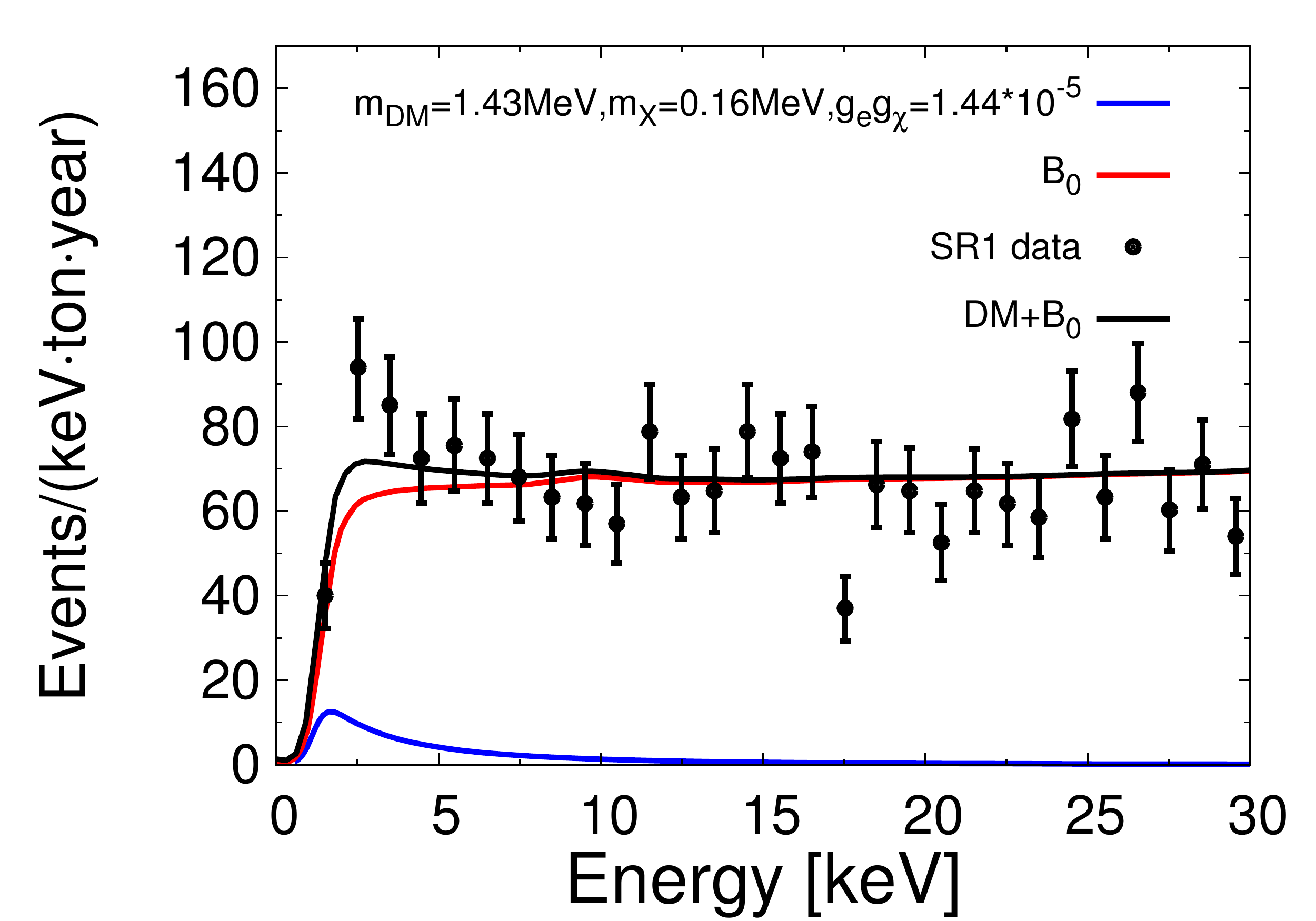}
\caption{\small \label{fig:best_fit}
[Scenario \#2: the $X$ gauge boson mediator case]
The full $\chi^2$ scanning over three parameters $(m_{\rm DM}, m_X, \sqrt{g_e g_\chi})$ is projected to $(m_X, \sqrt{g_e g_\chi})$, $(m_{\rm DM}, \sqrt{g_e g_\chi})$, and $(m_{\rm DM}, m_X)$ planes.
The best-fit point $(m_{\rm DM}, m_X, \sqrt{g_e g_\chi})=(1.43\,{\rm MeV}, 0.16\,{\rm MeV}, 3.79\times 10^{-3})$ is labelled with the ``star'' maker.
The bottom-right panel shows the event spectrum of the best-fit point with $\chi^2_{\rm Best-fit}=41.81$ and $p$-value=0.197.
}
\end{figure}

For the $X$ gauge boson mediator case, we have three independent parameters $(m_{\rm DM}, m_X, g_e g_{\chi})$ which are relevant to produce electron recoil spectra for XENON1T.
The spectral shape dependence on $m_X$ is depicted in Fig.~\ref{fig:DM_Xboson_XENON1T} for four benchmark values of $m_X=0.02, 0.1, 1$, and $10$ MeV with $m_{\rm DM}=0.1$ MeV.
Lighter $m_X$ provides narrower spectrum, and thus the XENON1T data prefers $m_X \lesssim 1$ MeV
in order to feature a peak around $2-3$ keV.
As can be seen in Fig.~\ref{fig:DM_Xboson_XENON1T_2}, the spectrum becomes narrower with larger $m_{\rm DM}$ in the range of $1\,{\rm keV} \lesssim m_{\rm DM} \lesssim 10\,{\rm MeV}$.
The spectral dependence observed in Figs.~\ref{fig:DM_Xboson_XENON1T} and~\ref{fig:DM_Xboson_XENON1T_2} is consistent with the mild correlation between $m_X$ and $m_{\rm DM}$ shown in the bottom-left panel of Fig.~\ref{fig:best_fit}.
The $\chi^2$ distributions scanned over three parameters $(m_{\rm DM}, m_X, g_e g_{\chi})$ are exhibited in Fig.\ref{fig:best_fit} for three different projection planes.
The linearly correlated region between $\sqrt{g_eg_\chi}$ and $m_X$ with $m_X \lesssim 2.5$ MeV is preferred by the XENON1T data, while the red region is excluded because of the over-production of the recoil events.
The minimum $\chi^2$ point, $(m_{\rm DM}, m_X, \sqrt{g_e g_\chi})=(1.43\,{\rm MeV}, 0.16\,{\rm MeV}, 3.79\times 10^{-3})$, labelled with ``star'' results in $\chi^2_{\rm best-fit}=41.81$, and the corresponding event spectrum is presented in the bottom-right panel of Fig.~\ref{fig:best_fit}.
The best-fit point gives $p$-value=0.197 compared to $B_0$, and thus the Scenario \#2 hypothesis is favoured by the data at $1.3\sigma$.

\begin{figure}[t]
\centering
\includegraphics[width=.8\textwidth]{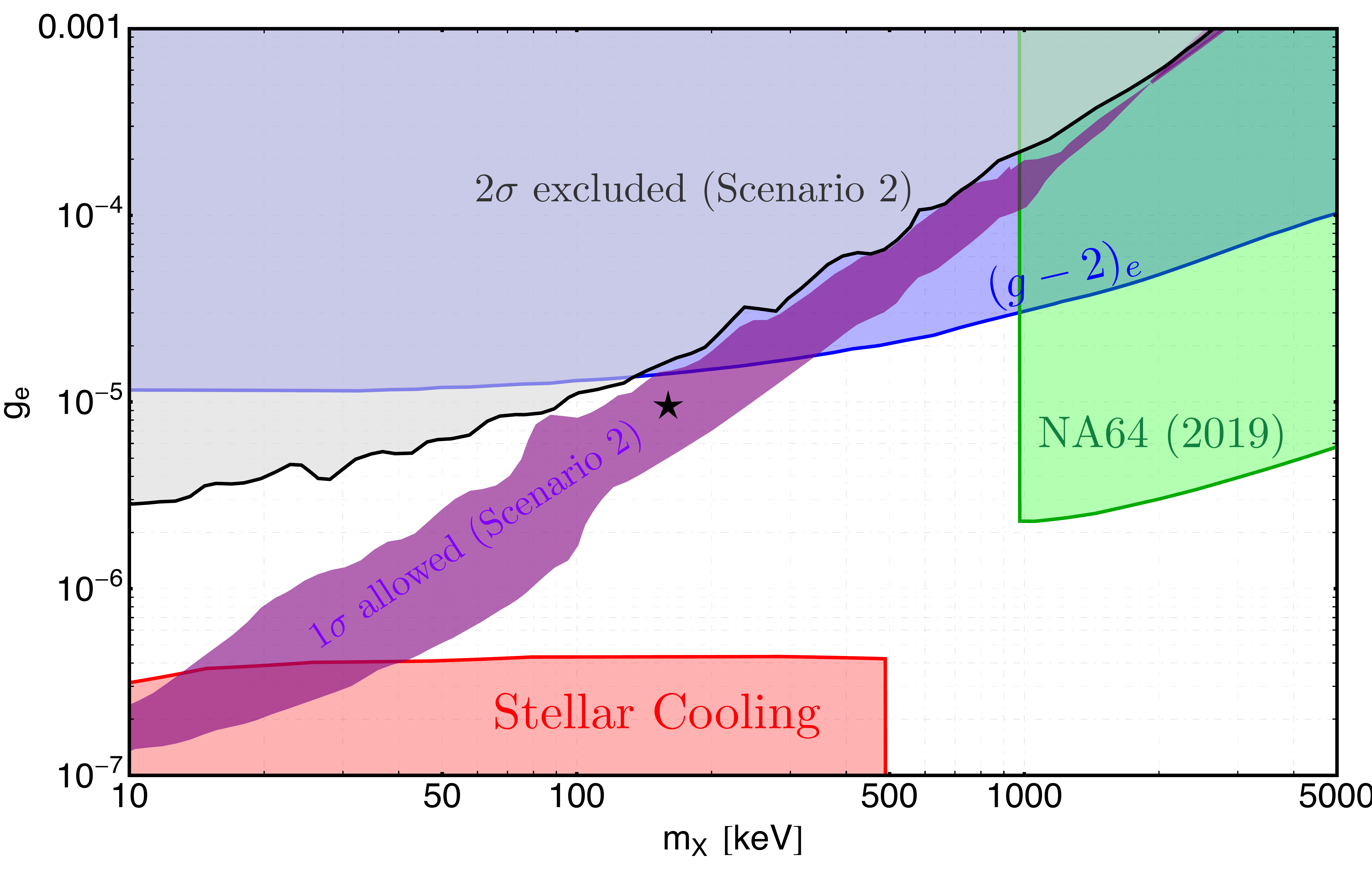}
\caption{\small \label{fig_scenario2_constraints}
[Scenario \#2] The favored (purple) and disfavored (gray) regions of the parameters ($m_X$, $g_e$) for $g_\chi = 1.5$ and $g_\nu = 0$.
The constraints on $g_e$ from $(g-2)_e$ (blue), stellar cooling (red), and NA64 (green) are also shown.
The ``star" marker indicates the same best-fit point as in Fig.~\ref{fig:best_fit}.
}
\end{figure}

We show other constraints on the mediator's electron coupling and mass for Scenario \#2 in Fig.~\ref{fig_scenario2_constraints} by fixing $g_\chi=1.5$ and $g_\nu=0$, and therefore the neutrino experiments and steller cooling through neutrino are not relevant.
Furthermore, missing $E_T$ searches via invisible decay of a dark gauge boson from NA64 show no constraint for the mass below some values because NA64 can just measure missing $E_T$ within their detection resolution.
Thus, for too light dark bosons ($m_X< E_{\rm resolution}$), they cannot provide any constraints.
Consequently, substantial of the XENON1T-preferred 1$\sigma$ region shown by purple-colored is still consistent with present observations.
See Section~\ref{sec:constraints} for more detailed explanation on other constraints.
As a method to prove the allowed region of Scenario \#2, we would like to discuss the process of neutron star heating by capturing the halo DM in Section~\ref{sec:NS}.

\section{Constraints from others} \label{sec:constraints}

The models in Scenario \#1 and Scenario \#2 can be probed by various searches as follows:
\begin{itemize}
\item $(g-2)_e$:
The anomalous magnetic moment of electron, $(g-2)_e$ can be enhanced due to the one-loop correction including a light $X$ boson.
For $m_X \ll m_e$, a sizeable coupling $g_e \gsim 10^{-5}$ is constrained~\cite{Pospelov:2008zw}.

\item \textit{Invisibly decaying dark photon search in NA64 experiment}:
For $m_X \gsim 1$ MeV, a large $X$ boson coupling to electron, $g_e \gsim 10^{-5} - 10^{-6}$, has been constrained by the dark photon search at NA64 experiment~\cite{NA64:2019imj}.

\item \textit{Stellar cooling constraints}:
If $X$ boson in the mass range $m_X \subset [0.1, 500]$ keV is weakly coupled to the electron, $g_e \subset [10^{-12}, 4\times 10^{-7}]$, $X$ boson can be produced inside the Sun and easily escape the Sun due to its long lifetime~\cite{Redondo:2008aa}.
In the absence of the neutrino coupling, $X$ boson with the coupling $g_e \gsim 4\times 10^{-7}$ is still allowed because it is captured by the Sun before the escape.
In the presence of neutrino coupling $g_\nu$, the produced $X$ boson predominantly decays into a neutrino pair.
The escape of these neutrinos from stellar objects will significantly increase the cooling rate, even with a larger coupling $g_\nu \gsim 4 \times 10^{-7}$~\cite{Harnik:2012ni, Davidson:2000hf}.

\item \textit{Neutrino-electron scattering}:
Borexino~\cite{Bellini:2011rx} and Gemma~\cite{Beda:2010hk} provide the limits on non-standard neutrino interaction by measuring $\nu-e$ scattering cross section, using $^7$Be solar neutrino and reactor neutrino, respectively.
We show their limits for the case of $g_e = g_\nu$ in Fig.~\ref{fig_scenario1_constraints}.

\end{itemize}

\section{ Neutron Star Heating} \label{sec:NS}

As an important prediction of our proposal in this paper is that a neutron star (NS) can be heated up to $\sim1500$ K by capturing DM particles in halo through DM-SM interactions and this will be tested by an infrared telescope in the near future, e.g., the James Webb Space Telescope~\cite{JWST}.

For Scenario \#2, the halo DM having coupling to electron can be captured by $~5\%$ electron component inside the neutron star~\cite{Bell:2019pyc}.
Before been captured, DM carries velocity larger than half speed of light due to the acceleration by the strong NS gravity.
After been captured, the DM kinematic energy transfers to the NS heat energy.
If the DM-electron cross section is large enough, the capture rate can reach the geometric limit~\cite{Garani:2018kkd}\footnote{The geometric limit implies that the entire halo DM in the vicinity of NS is captured.
The geometric limit of the NS capture rate is estimated by
\begin{equation}
C|_{\rm geom}= 5.6\times10^{25}
\left( \frac{\rho_\chi}{\rm GeV/cm^3}\cdot \frac{\rm 1 GeV}{m_{\rm DM}}
\cdot \frac{R_{\rm NS}}{\rm 11.6 km} \cdot \frac{M_{\rm NS}}{1.52 m_{\odot}}
 \right)~{\rm s^{-1}}\,,
\end{equation}
where $R_{\rm NS}$ and $M_{\rm NS}$ are the radius and mass of a typical NS star, respectively.
Rough estimation gives the critical cross section between DM and electron, $\sigma_{\rm DM-e}|_{\rm crit}\simeq \pi R^2_{\rm NS}/N_e\simeq 5 \times 10^{-44}\,{\rm cm^2}$.
}, such that the DM can heat up the NS and increase the NS temperature by $\sim1500$ K~\cite{Keung:2020teb}.
This $\sim1500$ K deviation of $\mathcal{O}(10^8)$ year old NS's temperature evolution will be sensitive to near future infrared telescopes.

The approximating capture rate by the electron component of NS is~\cite{McDermott:2011jp}
\begin{eqnarray}
\label{eq:Cc}
C_c\simeq \sqrt{\frac{6}{\pi}}\frac{\rho_{\rm DM}}{m_{\rm DM}}
\frac{v^2_{\rm esc}(R_{\rm NS})}{\bar{v}^2}
(\bar{v}\xi)
 N_e \int dK_e \frac{d\sigma_X({\rm DM}e \to {\rm DM}e)}{dK_e}\,,
\end{eqnarray}
which includes the Pauli blocking suppression factor $\xi\equiv \delta p/p_F$ where $\delta p$ is typical momentum transfer and $p_F$ is the Fermi momentum~\cite{McDermott:2011jp}.
Here $p_F\simeq 200$ MeV~\cite{Bell:2019pyc} and $\delta p\simeq \mathcal{O}({\rm keV})$ for $m_{\rm DM}\simeq \mathcal{O}({\rm keV})$.
Therefore, the Pauli blocking gives a suppression factor of $\xi\simeq \mathcal{O}(10^{-5}-10^{-6})$ for $\mathcal{O}({\rm keV})$ DM.
The escape velocity of the NS is $v_{\rm esc}(R_{\rm NS})=\sqrt{2GM_{\rm NS}/R_{\rm NS}}\simeq 0.63\,c$, the $\bar{v}$ is the DM dispersion velocity, the $\rho_{\rm DM}$ is the local DM density, and $N_e$ represents the total number of electrons in the NS.

For the effective DM-electron cross section from Fig.~\ref{fig:effective_sigma_XENON1T}, the best-fit point gives $\xi \sigma_{\rm DM-e}\simeq \mathcal{O}(10^{-37}-10^{-38})$, which is much larger than the critical cross section $\sigma_{\rm DM-e}|_{\rm crit}\simeq 5 \times 10^{-44}\,{\rm cm^2}$.
Therefore, the best-fit point and the entire region preferred by the XENON1T excess will be probed by infrared observations of the NS heating process in the near future.
For the $X$ gauge boson mediator case, substituting the best-fit value from Fig.~\ref{fig:best_fit} into Eq.(\ref{eq:Cc}) yields the capture rate $C_c\simeq 9.5\times 10^{37}~{\rm sec^{-1}}$ which is much larger than the geometric limit $C|_{\rm geom}\simeq 1.6\times 10^{28}~{\rm sec^{-1}}$, result in heating up the NS by $\sim1500$ K.
Even if we take a more extreme point within $1\sigma$ region, e.g., $(m_{\rm DM}, m_X, \sqrt{g_e g_\chi})=(0.1\,{\rm keV},0.03\,{\rm MeV},1\times 10^{-3})$, which is suppressed by $\xi\simeq \mathcal{O}(10^{-7})$, it yields the capture rate $C_c\simeq 4.3\times 10^{33}~{\rm sec^{-1}}$ which is still larger than the geometric limit $C|_{\rm geom}\simeq 2.2\times 10^{32}~{\rm sec^{-1}}$.

\section{Conclusion}

In this paper, we discussed the explanations of the recent XENON1T excessive electron recoil events around $2-3$ keV.
We considered a leptophilic vector mediator having couplings to leptons and dark matter in the keV-to-MeV mass range for two specific scenarios.
Both scenarios are simple anomaly free extensions of the standard model.

In Scenario \#1, the solar neutrino can easily generate the low-recoil energy excess for target electrons in XENON1T, in the presence of light mediator coupled to both electrons and neutrinos, i.e., $m_X \sim 10-100$ keV and $g_e = g_\nu \sim (3-5) \times 10^{-7}$.
Compared to the background model $B_0$, Scenario \#1 is favored by the data at $1.64\sigma$.
However, the stellar cooling and neutrino-electron scattering limits significantly constrain the favored region.

In Scenario \#2, we have studied the possibility to explain the XENON1T anomalous excess using boosted light dark matter, upscattered by energetic electron cosmic rays.
We first investigated the effective DM-electron scattering cross section case and found that light DM with $m_{\rm DM}\lesssim 10$ keV is preferred and there exists linear correlation between $\sigma_{\rm DM-e}$ and $m_{\rm DM}$ in the favored parameter space.
Next, we considered the $X$ gauge boson mediator to realize the DM-electron interaction and showed that the mediator mass $m_X\lesssim 2.5$ MeV is crucial to generate the peak around 2-3 keV in the electron recoil spectrum, and thus preferred by the XENON1T excess.
In the $1\sigma$ region, $\sqrt{g_e g_\chi}$ has strong correlation with mediator mass $m_X$ and mild correlation with DM mass in $1\,{\rm keV}\lesssim m_{\rm DM}\lesssim 10\,{\rm MeV}$.
Based on the $\chi^2$ minimization, $(m_{\rm DM},m_X,\sqrt{g_e g_\chi})=(1.43\,{\rm MeV},0.16\,{\rm MeV},3.79\times 10^{-3})$, with $\chi^2_{\rm best-fit}=41.81$, provides the best-fit to the data.
Then we studied other direct constraints on the electron coupling $g_e$ in the absence of the neutrino coupling ($g_\nu = 0$).
Requiring a large coupling between dark matter and mediator of $g_\chi \sim \mathcal{O}(1)$, we found that large portion of the favored region by the XENON1T anomaly is still allowed by the current constraints.
Finally, we emphasize that the observation of $\mathcal{O}(10^8)$ year old neutron stars by near-future infrared telescopes
will probe the allowed parameter region where the halo DM capture process by the electrons of NS is able to increase the NS temperature more than $\sim1500$ K.

\section*{Acknowledgments}
The work is supported in part by the National Research Foundation of Korea [NRF-2018R1A4A1025334, NRF-2019R1C1C1005073 (JCP), NRF-2019R1A2C1089334 (SCP), and NRF-2020R1I1A1A01066413 (PYT)].

\bibliography{refs}

\end{document}